\documentclass[12pt,preprint]{aastex}

\usepackage{amsmath}
\usepackage{natbib}
\shorttitle{\ion{He}{2} Ly$\beta$ Trough} 
\shortauthors{Syphers et al.}

\begin{document}


\title{\ion{He}{2} Ly$\beta$ Gunn-Peterson Absorption: New {\it HST} Observations, and Theoretical Expectations}

\author{David Syphers\altaffilmark{1},
Scott F. Anderson\altaffilmark{2},
Wei Zheng\altaffilmark{3},
Britton Smith\altaffilmark{4},
Matthew Pieri\altaffilmark{1},
Gerard A.\ Kriss\altaffilmark{3,5},
Avery Meiksin\altaffilmark{6},
Donald P.\ Schneider\altaffilmark{7},
J.\ Michael Shull\altaffilmark{1},
Donald G. York\altaffilmark{8,9}
}

\altaffiltext{1}{CASA, Department of Astrophysical and Planetary Sciences, University of Colorado, Boulder, CO 80309, USA; David.Syphers@colorado.edu}

\altaffiltext{2}{Astronomy Department, University of Washington, Seattle, WA 98195, USA}

\altaffiltext{3}{Department of Physics and Astronomy, Johns Hopkins University, Baltimore, MD 21218, USA}

\altaffiltext{4}{Department of Physics \& Astronomy, Michigan State University, East Lansing, MI 48824, USA}

\altaffiltext{5}{Space Telescope Science Institute, 3700 San Martin Drive, Baltimore, MD, 21218, USA}

\altaffiltext{6}{Scottish Universities Physics Alliance (SUPA), Institute for Astronomy, University of Edinburgh, Royal Observatory, Edinburgh EH9 3HJ, UK}

\altaffiltext{7}{Department of Physics \& Astronomy, Pennsylvania State University, 525 Davey Lab, University Park, PA 16802, USA}

\altaffiltext{8}{Department of Astronomy and Astrophysics, The University of Chicago, 5640 South Ellis Avenue, Chicago, IL 60637, USA}

\altaffiltext{9}{Enrico Fermi Institute, The University of Chicago, 5640 South Ellis Avenue, Chicago, IL 60637, USA}

\begin{abstract}
Observations of \ion{He}{2} Ly$\alpha$ Gunn-Peterson troughs have proved to be a valuable probe of the epoch of helium reionization at $z \sim 3$.
Since this optical depth can become unmeasurably large even for modest \ion{He}{2} fractions, various alternate techniques have been proposed to push to higher redshift, and among the more promising is looking at higher order Lyman-series troughs.
We here report four new observations of the \ion{He}{2} Ly$\beta$ trough, including new data on the only sightline with a prior Ly$\beta$ observation.
However, the effective optical depth ratio $\tau_{{\rm eff,}\beta}/\tau_{{\rm eff,}\alpha}$ is not simply predicted by $f_{\beta} \lambda_{\beta} / f_{\alpha} \lambda_{\alpha}=0.16$, and we analyze cosmological simulations to find that the correct ratio for helium at $z \sim 3$ is $\simeq$$0.35$.
In one case we infer $\tau_{{\rm eff,}\alpha} > 8.8$, strong evidence that helium was not fully reionized at $z=3.2$--$3.5$, in agreement with previous measurements suggesting a later completion of reionization.
\end{abstract}


\section{Introduction}
\label{sec:intro}

The full reionization of helium at $z \sim 3$ was a major step in the evolution of the intergalactic medium (IGM).
\ion{He}{2} requires four-Rydberg photons for ionization, and such hard photons are believed to be produced primarily in active galactic nuclei (AGN), rather than stars.
Thus the helium reionization epoch was delayed versus that of hydrogen (and \ion{He}{1}) at $z>6$ \citep[e.g.,][]{fan06}, occurring only when quasars became sufficiently numerous, probably beginning at $z \sim 3$--4 \citep[e.g.,][]{sokasian02,dixon09,syphers11}.
Helium reionization was not only a major change for the IGM state, but it also injected substantial heat into the IGM, and thus affected hydrogen as well.

Various indirect methods of constraining helium reionization via the \ion{H}{1} Ly$\alpha$ forest have been used, either by examining thermal broadening of Ly$\alpha$ forest features associated with the expected large heat input into the IGM \citep[e.g.,][]{ricotti00,schaye00,becker11} or by measuring changes in average Ly$\alpha$ forest opacity \citep[e.g.,][]{bernardi03,faucher-giguere08}.
The existence of the line-width change with redshift is subject to considerable controversy \citep{mcdonald01,kim02,meiksin10}, and while there appears to be a real opacity dip at $z \simeq 3.2$, its interpretation is not straightforward \citep{bolton09a,mcquinn09}.
Measurements of IGM metal line widths could break the thermal/non-thermal broadening degeneracy and yield firmer indirect constraints, but these are as yet observationally impossible in the necessary density regime \citep{meiksin10}.
Optical depth ratios comparing metal species whose ionization potentials straddle the \ion{He}{2} Lyman limit (54.4~eV, 228~\AA) may show a change in the UV background in $z \sim 3$--4 that is possibly associated with helium reionization \citep{songaila98,agafonova07}, but this observational result is disputed \citep{kim02a,aguirre04}, and such a change is not predicted in some models due to inhomogeneities in the UV background or in metallicity \citep{furlanetto09a,bolton11}.

The best constraints on helium reionization have come from direct \ion{He}{2} optical depth measurements, both at higher redshift \citep[$3.2 \lesssim z \lesssim 3.9$;][]{syphers11} and lower redshift \citep[$2.4 \lesssim z \lesssim 2.9$; e.g.,][]{shull10,worseck11a}.
However, there are regions of the Ly$\alpha$ Gunn-Peterson trough \citep{gunn65} with no detectable flux transmission in some individual spectra and multispectra composites.
Because the Gunn-Peterson optical depth of a transition depends on oscillator strength and wavelength, $\tau_i \propto f_i \lambda_i$, higher-order transitions have smaller opacities.
As a result, \ion{He}{2} fractions that are black in Ly$\alpha$ (giving only lower limits on the optical depth) can have measurable non-zero flux in higher-order troughs.
This method of looking at higher-order troughs has been used with \ion{H}{1} \citep[e.g.,][]{becker01} and proposed for \ion{He}{2} \citep{mcquinn09a}.
To date there has been only one \ion{He}{2} Ly$\beta$ optical depth measurement, for HE2347-4342 (henceforth HE2347; $z=2.9$), a challenging extraction of noisy data on the short-wavelength SiC detectors of {\it FUSE} \citep{zheng04b}.

In this paper, we compare \ion{He}{2}~Ly$\beta$ absorption, $\lambda_{\beta}=256.317$~\AA, to \ion{He}{2}~Ly$\alpha$, $\lambda_{\alpha}=303.782$~\AA, in three new sightlines, with data from the Cosmic Origins Spectrograph \citep[COS;][]{osterman11}.
We also analyze the COS data on Ly$\beta$ in the HE2347 sightline.
In Section~\ref{sec:obs} we describe the new observations of \ion{He}{2} quasars that allow Ly$\beta$ trough analysis, and in Section \ref{sec:tau_measurements} we discuss our optical depth measurements and the methodology used.

It is important to note that the simple relationship we have given above for optical depths of various transitions is valid for {\it effective} optical depths only under the assumption of a homogenous IGM, and does not hold in a clumpy IGM \citep{oh05}.
While this inhomogeneity effect has been taken into account in recent hydrogen Gunn-Peterson studies, it has been overlooked thus far in the sparse helium reionization literature dealing with higher order transitions.
In Section~\ref{sec:theory} we consider the relationship between effective $\tau_{\alpha}$ and $\tau_{\beta}$ in the context of analytical models and cosmological simulations.
In Section~\ref{sec:discussion} we discuss our new measurements in light of the opacity relationship established, and consider the implications for helium reionization.
We conclude in Section~\ref{sec:conclusion}.

\section{Observations}
\label{sec:obs}

In order to probe \ion{He}{2}~Ly$\beta$, quasars must have sufficient FUV flux, and also lie in a specific redshift range.
We need to observe the Ly$\beta$ Gunn-Peterson trough at a redshift where it is not completely obscured by a black Ly$\alpha$ Gunn-Peterson trough; in practice this means $z_{\alpha} \lesssim 2.8$, where the mean Ly$\alpha$ optical depth has dropped to $\sim$1--2.
This constraint requires that the quasars are at $z_{\rm QSO} \lesssim 3.5$.
At the other end, the effective area of COS drops sharply for $\lambda < 1150$~\AA, requiring the flux at the \ion{He}{2} break $f_{\lambda, \: {\rm break}} > 10^{-16}$~erg~s$^{-1}$~cm$^{-2}$~\AA$^{-1}$ for quasars at $z > 3.3$, and substantially larger at lower redshifts.
This is observationally challenging, as very few $z \sim 3$ quasars have any detectable FUV flux at all, although the lists of such \ion{He}{2} quasars have recently been rapidly expanding (\citealt{syphers09b,syphers09a,worseck11a}; Syphers et al., in prep).

Our sample in this paper consists of four \ion{He}{2} quasars observed with COS that fulfill these flux and redshift requirements.
The first was observed in an ongoing campaign to verify more \ion{He}{2} quasars, {\it HST} program GO~12178 (Syphers et al., in prep.), with a relatively short reconnaissance spectrum: SDSSJ0915+4756 (henceforth SDSS0915, $z=3.34$).
Two more were observed in a followup program, {\it HST} GO~12249 (Zheng et al., in prep.), with long exposures: SDSSJ1253+6817 (henceforth SDSS1253, $z=3.47$), and SDSSJ2346-0016 (henceforth SDSS2346, $z=3.51$).
SDSS0915 is newly verified as a \ion{He}{2} quasar, while SDSS1253 was verified in \citealt{syphers09b} and SDSS2346 in \citealt{zheng04a}.
Below 1100~\AA\ the COS effective area is very small, so for quasars at lower redshifts (down to $z \sim 2.7$), the required flux to measure Ly$\beta$ Gunn-Peterson is $f_{\lambda, \: {\rm break}} \gtrsim $ few$\;\times 10^{-15}$~erg~s$^{-1}$~cm$^{-2}$~\AA$^{-1}$.
Currently, HE2347 \citep{shull10} uniquely satisfies this requirement, and completes our present sample.

We observed our new targets with COS/G140L ($R \sim 2000$--3000), and adopt 7 pixels ($0.56$~\AA) as a resolution element.
This is approximately the FWHM of the line-spread function (LSF), although the LSF is non-Gaussian with substantial wings, due primarily to mirror polishing errors on {\it HST} \citep{kriss11}.
The data were reduced using CALCOS 2.13.6, and we coadded individual exposures with custom software that flatfields the data and takes into account various detector defects.
For further details on the GO 12178 observations and the removal of geocoronal emission lines, see Syphers et al.\ (in prep.), and on the 12249 observations, see Zheng et al.\ (in prep.).
We are interested in broad effective optical depth measurements with flux averaged over all available trough regions, but in any case, averaging over substantial optical depth fluctuations is unavoidable for G140L data, because single resolution elements have widths of $\sim$100--160~km~s$^{-1}$.
HE2347 was observed with COS/G130M ($R \sim 16$,000--20,000) and COS/G140L; for details on that observation, see \citet{shull10}.
Since G130M currently covers only $\lambda > 1134$~\AA, we use this data only for the Ly$\alpha$ opacities, and use G140L for Ly$\beta$.
The observations are summarized in Table~\ref{tab:obs}.

HE2347 was previously observed with {\it FUSE} for 619~ks \citep{kriss01,zheng04b}.
COS/G140L has $\sim$10~cm$^2$ effective area at very short wavelengths \citep[$\lambda \lesssim 1000$~\AA;][]{mccandliss10}, which is similar to or slightly higher than {\it FUSE} \citep{sahnow00}.
For HE2347, these G140L data constitute only $\sim$12~ks of the COS observations \citep{shull10}, and there would therefore appear to be little hope of improving upon the {\it FUSE} measurement.
However, the {\it FUSE} detector consisted of four different channels of two segments each, with two different coatings, and it often proved challenging to match flux calibration across segments, particularly for the lower-wavelength SiC channels.
In addition, {\it FUSE} had a much larger stray and scattered-light background than COS, and although the dark current was about the same as COS per 2-D pixel, it was on average nearly 50 times larger than COS/G140L per Angstrom in the 1-D spectrum.
We thus deem it worthwhile to analyze the COS HE2347 data to see what limits they set on $\tau_{\beta}$.

\begin{figure}
\epsscale{1.0}
\plotone{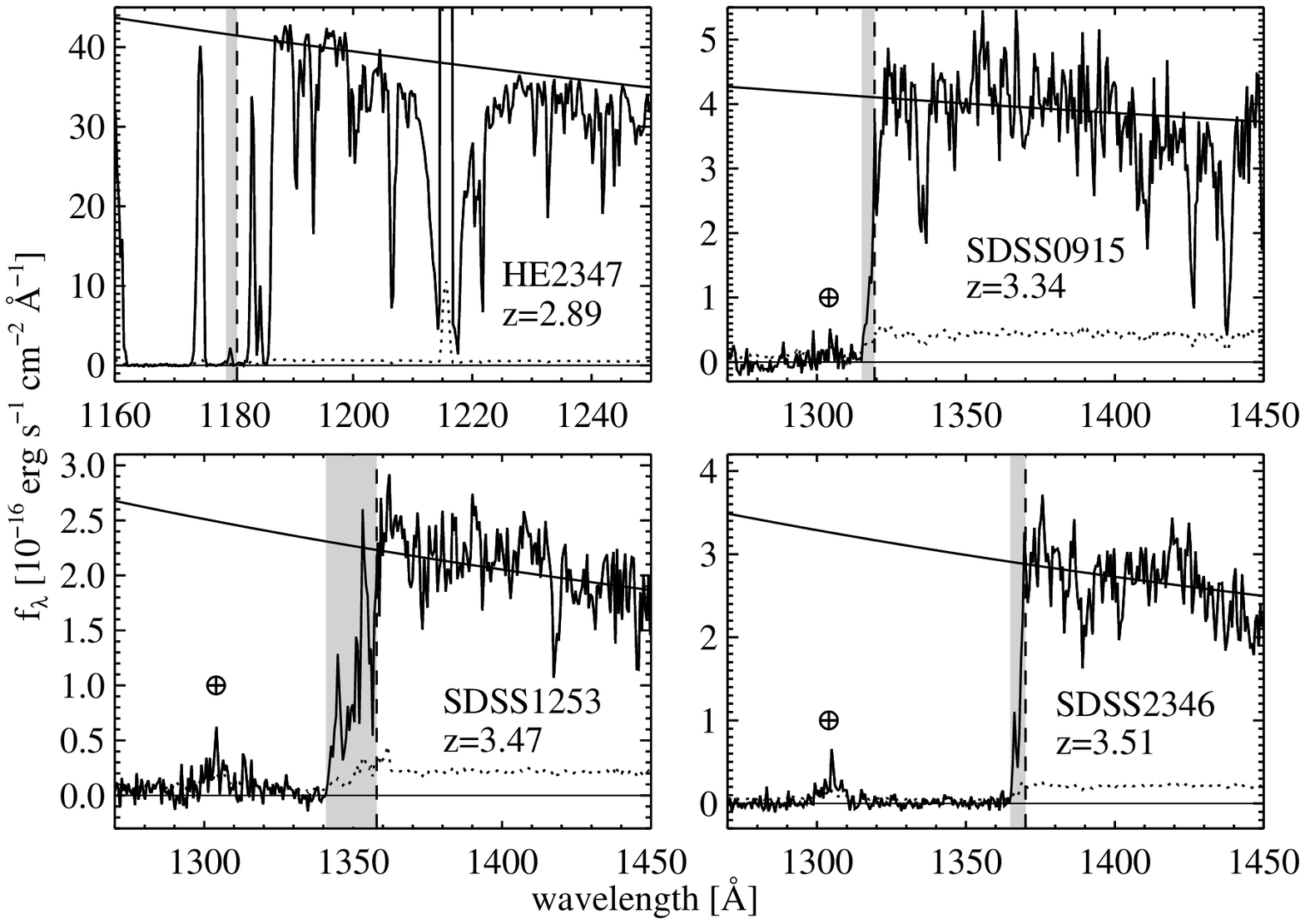}
\caption{Regions near \ion{He}{2}~Ly$\alpha$ for the four quasars studied in this paper. The vertical dashed line indicates the onset of \ion{He}{2}~Ly$\alpha$ absorption according to the quasar redshift, and the shaded region is the ionized proximity zone, identified in Ly$\alpha$. The redshifts covered by the proximity zone are excluded from Ly$\beta$ analysis. Continuum fits are overplotted (solid lines), and use data from the \ion{He}{2}~Ly$\alpha$ break to a S/N cutoff in the red (typically near 1900~\AA). Residual geocoronal \ion{O}{1} emission is visible near 1302~\AA\ (marked).}
\label{fig:lya_spec}
\end{figure}

For the three new G140L observations, we fit the quasar continua with power laws over the entire available wavelength range, from the \ion{He}{2}~Ly$\alpha$ break to a S/N cutoff in the red (typically near 1900~\AA).
We first deredden the spectra with a Fitzpatrick \& Massa UV extinction curve \citep{fitzpatrick99}, using $E(B-V)$ from \citet{schlegel98}.
For the fit, we specify regions that appear to be largely free of \ion{H}{1}~Ly$\alpha$ absorption to initialize, and then iteratively use all consistent continuum points.
HE2347 has much higher-resolution G130M data, as well as much better S/N, which allows the fit regions to be defined quite precisely.
The G130M data also allow a careful fit of Galactic \ion{H}{1}~Ly$\alpha$ absorption, and thus an $E(B-V)$.
We therefore use the power law index and normalization, and the extinction, from \citet{shull10} for this object.
The continuum fits are shown in Figure~\ref{fig:lya_spec}, along with the quasar proximity zones, which we exclude from our optical depth calculations.

\section{Optical Depth Measurements}
\label{sec:tau_measurements}

Our goal of estimating \ion{He}{2}~Ly$\alpha$ effective optical depths is reached in three steps: (1) we measure a raw \ion{He}{2}~Ly$\beta$ $\tau_{\rm eff}$ directly from each spectrum, (2) we correct this measurement for contaminating lower-redshift \ion{He}{2}~Ly$\alpha$ $\tau_{\rm eff}$ by one of three different methods, and (3) we predict $\tau_{{\rm eff,}\alpha}$ using our corrected $\tau_{{\rm eff,}\beta}$ and a conversion factor found from cosmological simulations.
We describe the first two steps in this section, and the last step in section \ref{sec:theory}.

\subsection{Optical Depth Methodology}
\label{sec:tau_methods}

Calculating optical depths in very low-count regimes, such as black troughs or low-sensitivity portions of the detector, requires care.
We here set forth a methodology appropriate for such calculations, which not only gives good results for individual spectra, but enables comparison between different spectra.

The COS detector has a very low background that is dominated by dark current ($\sim$$10^{-4}$~counts~s$^{-1}$~pix$^{-1}$, for a pixel in the 1-D spectrum).
Nonetheless, in Gunn-Peterson troughs with near-zero flux, this background can be substantially larger than the signal.
We are thus measuring signal counts, $n_s$, and background counts, $n_b$, in a regime where we can have both $n_s \ll n_b$ and $n_s \sim 0$.
The CALCOS pipeline takes $n_s = n_{\rm obs} - \langle n_b \rangle$ in each pixel, where the average background is found from regions offset in the cross-dispersion direction to unexposed portions of the detector, and smoothed over 100 pixels.
Because the low background means that often $n_{\rm obs} < \langle n_b \rangle$ (or indeed $n_{\rm obs}=0$) in any given pixel, we often have the unphysical result that $n_s < 0$.
Thus the nonparametric bootstrap median method we developed for use with ACS prism data \citep{syphers11} is not useful here.

Even assuming a simple parametrization (e.g., Poissonian) does not lead to one clearly correct way of determining confidence intervals when near a physical boundary \citep{mandelkern02}.
However, the most important thing is to choose a consistent method for analyzing all relevant observations and simulated spectra, and for this we adopt the ``unified'' frequentist confidence intervals of \citet{feldman98}, which use likelihood ratio ordering.
This is a standard in the particle physics community when working near physical boundaries, and in our simulations of COS spectra, it generally has acceptable coverage even for very high optical depths.
Feldman-Cousins confidence intervals also have the desirable property that they are ``unified'', i.e., as the signal is reduced compared to the background and zero signal becomes included in the lower limit, the intervals naturally transition from two-sided to one-sided intervals with the same coverage.
For classical Neyman confidence intervals, if the decision whether or not to quote one-sided or two-sided intervals is based on the data (as it typically is in astronomy; \citealt{feldman98} call this ``flip-flopping''), then the coverage from repeated experiments is {\it not} the claimed coverage.

One important aspect to note is that because Feldman-Cousins confidence intervals are unified, they are not central (that is, repeated measurements yield confidence intervals which do not necessarily have equal probability of lying above or below the true parameter value).
They do transition into central intervals for large signal counts, but we caution that when we deal with high optical depths, we are of course in the low-count regime.

For troughs where the total signal count is formally negative, we recommend quoting two values in addition to the confidence interval.
First, the ``sensitivity'' of the experiment, as defined by \citet{feldman98}, which is the average upper limit obtained for many experiments with the given Poissonian background but zero signal.
Second, similarly motivated, the ``detection upper limit,'' as defined by \citet{kashyap10}\footnote{Note that \citet{kashyap10} use the terms ``upper bound'' and ``upper limit'' to distinguish between quantities that we prefer to call, more descriptively, ``source upper limit'' and ``detection upper limit''. The latter is a property of the detector, and does not depend on the source examined.}.
As with the sensitivity, the detection upper limit depends purely on the detector, rather than the source, and is defined as the maximum intensity a source can have without having at least a probability $\beta$ of being detected at a significance level $\alpha$.
Note that we use frequentist Feldman-Cousins confidence intervals for both calculations, rather than Bayesian credible intervals as used by \citet{kashyap10} for the latter quantity.

We estimate the expected number of counts, absent any absorption, by extrapolating the quasar continuum fit into the Gunn-Peterson trough, and multiplying by the flatfield.
The primary flatfield features are shadows made by the grid wires above the detector \citep{osterman11}, which reduce the number of counts seen in those regions.
Multiplying by the flatfield puts these shadows into the extrapolated quasar continuum, similarly reducing the expected counts.
We also take care to use the time-dependent sensitivity function to convert fluxes into counts for a given observation, since COS sensitivity does change over time \citep{osterman11}.
We derive expected counts using the exact same detector regions as for the actual source counts and background for each individual exposure, which enables direct comparison between the two.
Source and background counts from all relevant exposures are combined to derive a source counts confidence interval, which is converted to an optical depth confidence interval using the expected number of counts.

\subsection{Removing \ion{He}{2}~Ly$\alpha$ Contamination}
\label{sec:removing_contamination}

Our targets were chosen at redshifts high enough so that the Ly$\beta$ trough is accessible to COS, but low enough so that it lies in the \ion{He}{2}~Ly$\alpha$ forest, rather than the Ly$\alpha$ Gunn-Peterson trough.
Nonetheless, the opacity in the helium Ly$\alpha$ forest is substantial at these redshifts ($\sim$1--2), and must be taken into account.
We consider three different methods to do so, each with systematics that are large, but at least different from each other.

Method 1 (M1) is the technique commonly used when calculating \ion{H}{1}~Ly$\beta$ Gunn-Peterson opacity, which is to remove the mean Ly$\alpha$ opacity at the given redshift, using a fit calculated from spectra of other, lower-redshift quasars \citep[e.g.,][]{becker01}.
There have been such fits of mean \ion{He}{2} opacity, but based only on a few sightlines (no more than two at any given redshift), and they are thus subject to large systematic uncertainty, as well as simply having very large scatter \citep{shull10,worseck11a}.
In addition, the effective optical depths used in the fits have often been calculated for selected coherent regions of high or low $\tau$, rather than averaging over redshift bins without regard for chance structure, which distorts the results.
Of course, the substantial variance between sightlines also implies that the average opacity, even perfectly determined, may not correspond well to the actual opacity in a specific sightline.
We use the simple semianalytical $\eta=80$ model of \citet{worseck11a} to estimate the average $\tau_{{\rm eff,}\alpha}(z)$, as it provides a good fit to the optical depths of three \ion{He}{2} quasars in the Ly$\alpha$ forest at $z=2.3$--$3.0$.
(This model is not a direct fit of the \ion{He}{2}~Ly$\alpha$ data, relying as it does on a hydrogen distribution and $\eta \equiv N_{\rm He \, II}/N_{\rm H \, I}$, but it nonetheless fits the data quite well, and allows sensible extrapolation beyond where we have data---namely down to $z=2.1$, where we need the opacity for the HE2347 sightline.)

The accuracy of this model over small wavelength ranges in our data can be tested by comparing the M2 Ly$\alpha$ measurements (free of Ly$\beta$; see below) to model predictions over the same redshift range.
The model underpredicts the opacity for the higher-redshift quasars, and overpredicts the opacity for HE2347, but the differences are relatively small compared to the opacity change seen in and out of the Ly$\beta$ trough.
For HE2347, SDSS0915, SDSS1253, and SDSS2346, respectively, the difference between the value measured just redward of the Ly$\beta$ trough and the model prediction for that region is $\Delta \tau = -0.30$, $0.30$, $0.47$, and $0.75$.
Such differences are of the magnitude expected given sightline variations in density and ionizing background (and thus $\eta$), but demonstrate why this method is not perfect, even if the average opacity model were derived from more than a few quasars.

Method 2 (M2) uses the opacity of the \ion{He}{2}~Ly$\alpha$ forest just redward of the Ly$\beta$ trough as an estimate for Ly$\alpha$ contamination in the trough.
This approach relies on the effective Ly$\alpha$ optical depth not changing substantially over small redshift intervals ($\Delta z \sim 0.05$--$0.1$), which is clearly not a perfect assumption.
It has the advantage over the first method that we are comparing two regions of the same sightline, where we know reionization has completed (from an observational perspective, $\tau \lesssim 5$), as opposed to comparing different sightlines, which can complete reionization at somewhat different redshifts.
On the other hand, it should tend to give a somewhat higher Ly$\alpha$ opacity than the true value (because the opacity decreases with redshift on average), which would underestimate the Ly$\beta$ opacity.
In practice, we use the 20~\AA\ above the Ly$\beta$ break to calculate the effective optical depth.

Method 3 (M3) exploits a unique possibility when analyzing helium, not available to hydrogen Gunn-Peterson studies.
Because we are working with FUV spectra in helium, we can use the optical spectrum to probe the same redshift in hydrogen, allowing us to see and account for IGM structure.
For HE2347, we use the VLT/UVES spectrum \citep{kim07}, and for the other quasars, we use spectra from the Sloan Digital Sky Survey \citep[SDSS;][]{york00}.
This may appear to be the best method, since it is the only one to give information on the very same IGM region that is causing our \ion{He}{2}~Ly$\alpha$ absorption.
Nonetheless, it too has notable drawbacks in practice.
It relies on converting \ion{H}{1} opacity to \ion{He}{2}, $\tau_{{\rm He \, II}}/\tau_{{\rm H \, I}} \simeq \eta/4$ (where the factor of 4 comes from the wavelength dependence).
Typical values are $\eta \sim 50$--100, but there is substantial variation in $\eta$ throughout the IGM, and this changes during reionization \citep{zheng04b,fechner06,shull10}.
As a result, this is our primary uncertainty in this method.
This conversion becomes more complicated when the optical spectra do not resolve IGM structure, as is the case with our new targets and their SDSS spectra ($R \sim 1800$ in our region of interest), although not with the VLT spectrum of HE2347.

\begin{figure}
\epsscale{1.0}
\plotone{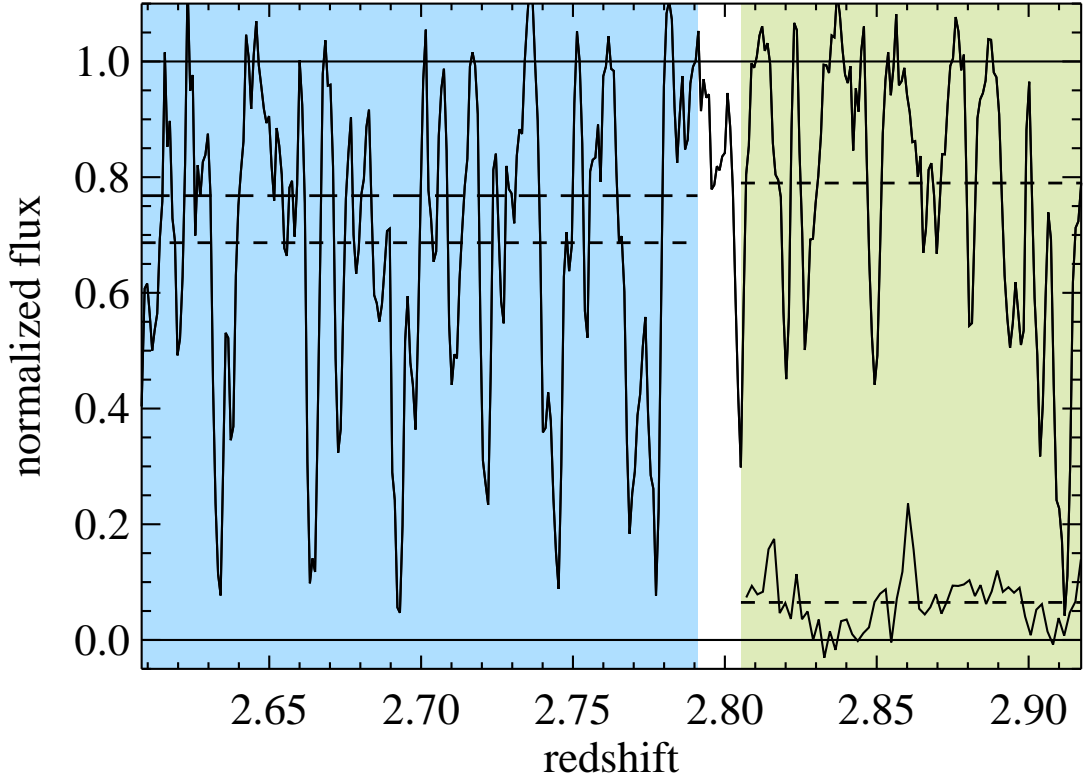}
\caption{Continuum-normalized spectrum of SDSS2346, as an example of how the M3 estimator is derived. The upper curve is the optical data for \ion{H}{1}, and the lower curve (shown only for $z>2.805$) is the UV data for \ion{He}{2}. Solid horizontal lines indicate fluxes of 0 and 1, while short-dashed horizontal lines indicate the mean measured flux for the given spectrum in the relevant shaded region. $n_{\rm eff}$ is found by comparing these mean fluxes in the calibration region (the right shaded region; green in the online version), where there is only Ly$\alpha$ absorption. The mean flux in the region where M3 is measured (the left shaded region; blue in the online version) is contaminated by \ion{H}{1}~Ly$\beta$ absorption from higher redshift; when we correct for this, the mean flux moves up to the long-dash line.}
\label{fig:HI_regions}
\end{figure}

We calibrate our effective $\eta_{{\rm eff}} = 4 \times (\tau_{{\rm eff, \: He \, II}})/(\tau_{{\rm eff, \: H \, I}})$ by considering redshift regions of the \ion{He}{2}~Ly$\alpha$ forest that are not Gunn-Peterson opaque, but are redward of the onset of Ly$\beta$ absorption (see Figure \ref{fig:HI_regions}).
In these regions we can measure the effective Ly$\alpha$ opacities with no Ly$\beta$ contamination (in practice, we also avoid some regions of poor continuum fitting in the optical data near the \ion{O}{6} emission line).
In all three sightlines for which we have SDSS data, this comparison gives similar $\eta_{{\rm eff}}$, and we thus adopt the single value $\eta_{{\rm eff}} \simeq 44 \pm 5$ for all our sightlines.
This approach neglects the redshift evolution of $\eta$, as well as continuum uncertainties, which are hard to characterize for the SDSS data.
There may be a weak trend of increasing $\eta_{{\rm eff}}$ with increasing redshift, but we caution that we are not measuring $\eta$, which is defined point by point rather than averaged over a large region like $\eta_{{\rm eff}}$.
So although our SDSS $\eta_{{\rm eff}}$ is consistent with, e.g., the mean $\langle \eta \rangle \simeq 50$ found at lower redshift in the HE2347 sightline \citep{shull10}, this probably has little significance.
Indeed, for HE2347 itself we prefer $\eta_{{\rm eff}} = 26 \pm 5$, and use this value with the VLT data.

An additional source of uncertainty in this method is continuum fitting the SDSS data, for which we use a well-established spline-fitting routine.
We tested this routine on mock SDSS spectra of comparable S/N to our actual quasars, and found that it introduced errors on average $\simeq$10\% (ranging from 1\% to 20\%) in the \ion{H}{1}~Ly$\alpha$ $\tau_{\rm eff}$ measurement, typically underestimating the opacity.
This is a non-trivial source of error, but the systematic uncertainties in $\eta_{{\rm eff}}$ dominate the error budget.
The presence of \ion{H}{1}~Ly$\beta$\ and \ion{O}{6} quasar emission do complicate the optical spectrum, but the redshift region of the Ly$\beta$ trough is largely unaffected, due to our exclusion of any proximity zone (see Figure \ref{fig:lya_spec}).

Before converting $\tau_{{\rm eff, \: H \, I}}$ into $\tau_{{\rm eff, \: He \, II}}$, there is a small correction we take into account.
Because this region of the \ion{H}{1}~Ly$\alpha$ forest also contains higher-redshift \ion{H}{1}~Ly$\beta$ lines as well, we remove that source of opacity.
To do this, we measure the relevant \ion{H}{1}~Ly$\alpha$ forest at higher redshift, and estimate the concomitant \ion{H}{1}~Ly$\beta$\ forest, using mean flux decrements and the effective \ion{H}{1} opacity ratio $\tau_{{\rm eff,}\beta}/\tau_{{\rm eff,}\alpha}=0.29$ (as found in Section \ref{sec:theory}).
The effect of this is shown in Figure \ref{fig:HI_regions}.
It might initially appear interesting to examine the \ion{He}{2}~Ly$\beta$ opacity {\it without} this last step, because using all \ion{H}{1} opacity (including Ly$\beta$) when converting to helium opacity should give a \ion{He}{2}~Ly$\beta$ optical depth from only the diffuse IGM not detected in \ion{H}{1} absorption.
This also being opaque would be very strong evidence for Gunn-Peterson absorption, but in practice, the uncertainty in $\eta_{\rm eff}$ makes such a test impractical.

\subsection{Optical Depth Results}

\begin{figure}
\epsscale{1.0}
\plotone{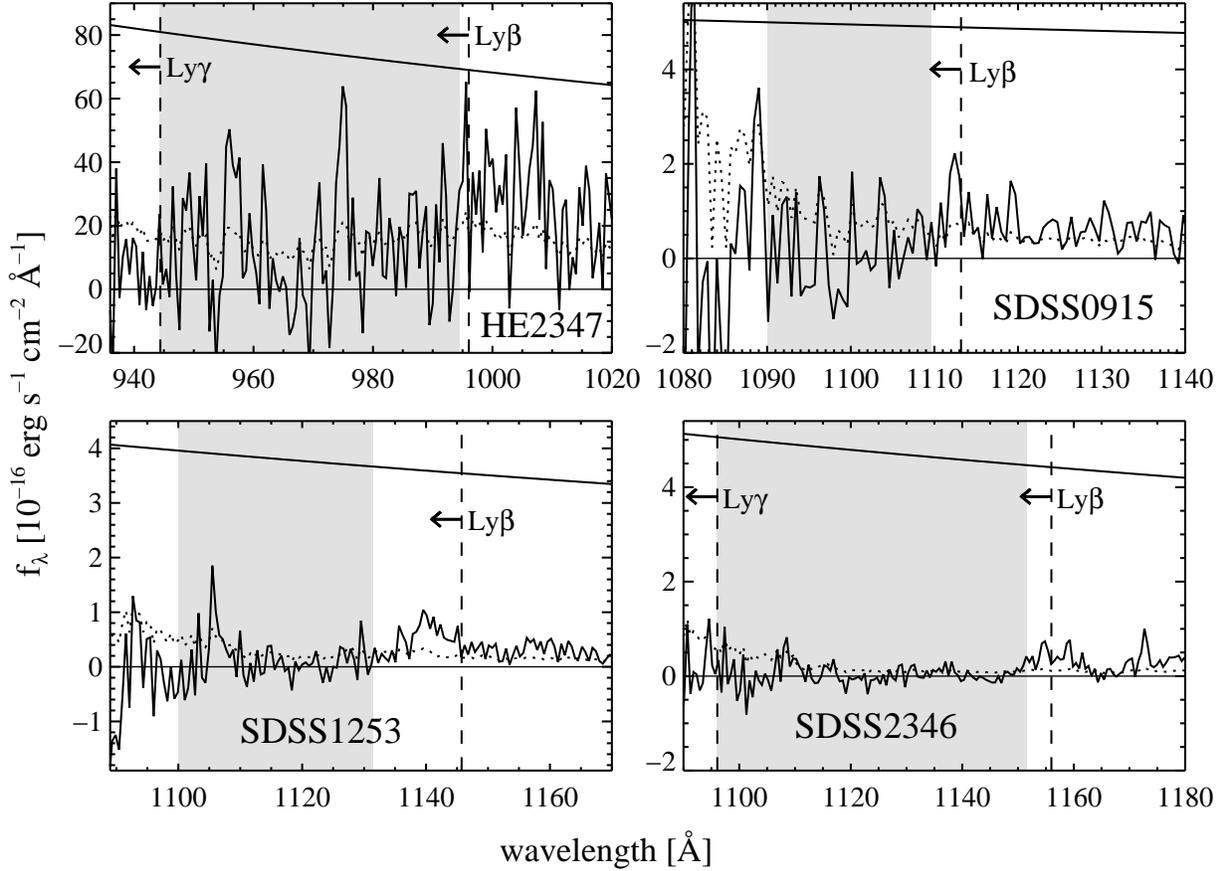}
\caption{\ion{He}{2}~Ly$\beta$ regions of the four quasars used in this paper. Extrapolated continuum fits are overplotted (solid lines). The vertical dashed lines indicate the onset of \ion{He}{2}~Ly$\beta$ or Ly$\gamma$ absorption, and the shaded region indicates the Ly$\beta$ trough as extracted. This trough avoids the proximity region identified in Figure \ref{fig:lya_spec}. The blue edge of the trough is set by the onset of the Ly$\gamma$ trough in HE2347 and SDSS2346, and by very low-S/N regions in SDSS0915 and SDSS1253. Ly$\beta$ optical depths for the latter two are robust under $\sim$10~\AA\ changes in the blue edge, because the sensitivity is so low there.}
\label{fig:lyb_spec}
\end{figure}

For each object, we exclude any possible proximity zone (as seen in \ion{He}{2}~Ly$\alpha$; see Figure \ref{fig:lya_spec}), where radiation from the quasar itself noticeably affects local optical depths.
We then measure the opacity of the Ly$\beta$ troughs (Figure \ref{fig:lyb_spec}), using the methodology of Section \ref{sec:tau_methods}.

The size of the Ly$\beta$ trough we can measure is limited by two factors.
For HE2347 and SDSS2346, we use the full trough from the onset of Ly$\beta$ absorption down to the onset of Ly$\gamma$ absorption ($243.027$~\AA\ rest frame).
For SDSS1253 and SDSS0915, our lower limits are determined by S/N, itself set by the rapidly declining response of the COS detector towards shorter wavelengths.
(We can test the difference of these two cutoffs in SDSS1253, where we have data of quality that is usable, albeit low, down to the Ly$\gamma$ break.
Using this larger definition of the trough yields results very consistent with the S/N cut, although with larger error bars due to inclusion of very low-S/N data.)

We remove geocoronal emission lines by using only data taken during orbital night in those regions affected (for details, see Syphers et al., in prep.)
Despite this removal, a remaining concern might be that the very high flux of the geocoronal lines (particularly Ly$\alpha$) is large enough that scattered light from the lines will contaminate the spectrum far from the line wavelengths.
To test this possibility, we calculate optical depths in troughs two ways, first using all data taken, and then using night-only data.
(Although geocoronal Ly$\alpha$ is present at all times, it is substantially weaker at night.)
In every case the results are consistent, including for the highest optical depths, and thus scattered geocoronal emission is negligible.

Results of our measurements are presented in Table~\ref{tab:obs_opacities}.
The columns of this table are as follows.
Column~(1)---target name.
Column~(2)---target redshift, taken from a combination of the onset of IGM \ion{He}{2}~Ly$\alpha$ absorption and rest-frame UV emission lines seen in the optical, except for HE2347, for which the redshift is taken from [\ion{O}{3}]~$\lambda$5007 (R.\ Simcoe, personal communication), and which agrees with \ion{O}{1}~$\lambda$1302 \citep{reimers97}.
Column~(3)---the redshift range used for the \ion{He}{2}~Ly$\beta$ trough (Fig.\ \ref{fig:lyb_spec}), avoiding any quasar proximity zone, the Ly$\gamma$ trough, and extremely low-S/N areas.
Column~(4)---the measured \ion{He}{2}~Ly$\beta$ trough optical depth, with no correction for contaminating Ly$\alpha$ opacity. Quoted errors are 68\% Feldman-Cousins confidence intervals.
Columns~(5--7)---estimated optical depth for \ion{He}{2}~Ly$\alpha$ in this redshift region, using methods M1, M2, and M3, described in Section \ref{sec:removing_contamination}.
Errors in column~7 are 68\% random only.

\section{Predicted Optical Depth Ratios}
\label{sec:theory}

It is widely understood that $\langle \tau \rangle = \langle -\ln{(F/F_0)} \rangle$ does not equal $\tau_{{\rm eff}} = -\ln{(\langle F/F_0 \rangle)}$ in an inhomogeneous IGM \citep[e.g.,][]{dixon09}, but less widely appreciated is the implication that $\tau_{\alpha}/\tau_{\beta} \ne \tau_{{\rm eff,} \alpha}/\tau_{{\rm eff,} \beta}$.
As a result, when using $\tau_{{\rm eff,} \beta}$ to estimate $\tau_{{\rm eff,} \alpha}$, we cannot simply use the ratio of oscillator strength and wavelength ($f_i \lambda_i$), as has been done in the heretofore sparse \ion{He}{2} literature on this topic, as well as early \ion{H}{1} Gunn-Peterson studies.
\citet{oh05} pointed out this issue, and derived a ratio applicable for $z \sim 6$ hydrogen, but we need to reassess this for our $z \sim 3$ helium troughs.

To characterize this effective optical depth ratio, we turn to the cosmological simulations of \citet{smith11}.
These simulations use the Eulerian hydrodynamics $+$ $N$-body code \texttt{Enzo}, with adaptive mesh refinement turned off to allow uniformly good resolution in the low-density IGM.
The simulation box was 50~$h^{-1}$~comoving~Mpc on a side, with $1024^3$ cells, a dark matter particle mass of $7 \times 10^6 \; h^{-1}$~$M_{\sun}$, and distributed feedback \citep[simulation 50\_1024\_2 of ][]{smith11}.
Radiation post-processing for metal species was done with the UV background models of \citet{haardt01}, assuming equilibrium; non-equilibrium ionization states of H and He were calculated as the simulation ran.
For details on the feedback method, the ionization calculation, and other aspects, see \citet{smith11}.

We take 500 rays through this volume at each of several redshifts from $z=2.5$--$3.8$, and create a high-resolution ($R > 50$,000) simulated spectrum for each ray.
We verify that point-by-point, $\langle \tau_{\beta}(\lambda)/\tau_{\alpha}(\lambda) \rangle = 0.160$ and $\langle \tau_{\gamma}(\lambda)/\tau_{\alpha}(\lambda) \rangle = 0.0557$, showing that we have enough resolution to recover the $f_i \lambda_i$ ratio we expect in this case (Table~\ref{tab:theory_opacities}).

To test how these spectra reproduce the $z \sim 3$ forest, we compare the hydrogen $\tau_{{\rm eff,} \alpha}(z)$ to the observational values from \citet{faucher-giguere08}.
The simulations track the effective optical depth quite well, with the exception of the highest-redshift simulation point used ($z \simeq 3.7$), which differs by $0.3$~dex from the observed value, and a $\sim$2$\sigma$ discrepancy near the $z \sim 3.2$--$3.3$ opacity dip.
This apparent observed dip (a deviation from a power law fit) is not well understood, nor well reproduced by simulations \citep[e.g.,][]{faucher-giguere08,bolton09a}.
However, the opacity ratio of interest to us is fairly insensitive to minor disagreements such as this; over the redshift range $z=2.5$--$3.7$, our hydrogen and helium optical depths change by a factor of $\sim$4, but the ratio $\tau_{\beta}/\tau_{\alpha}$ varies by only 7\% for helium.
We also compared the flux probability distribution function of the simulation with the observed values from \citet{kim07}, at $z=3.0$.
There is good agreement here except in the two highest flux bins (transmission $T > 0.925$), where some substantial discrepancy is predicted based on problems finding the continuum observationally at higher redshift \citep{faucher-giguere08}.

The effective optical depth ratios predicted by the simulation, contrasted with the naive expectation, are shown in Table~\ref{tab:theory_opacities}, the columns of which are as follows.
Column~1---transition.
Column~2---wavelength of the transition \citep{ralchenko08}.
Column~3---oscillator strength calculated from equation~8 of \citet{meiksin09} (which agrees with \citealt{verner96} to the precision that the latter reference quotes).
Column~4---the predicted optical depth ratio given by the ratio of $f_i \lambda_i$.
Column~5---the ratio actually seen in simulations, for redshift $z \simeq 3.3$.

The optical depth ratios do evolve with redshift, albeit weakly, trending towards the point-by-point value as the mean opacity drops.
We find for \ion{He}{2} at redshifts $z=2.4$--$3.8$, to within 2\% for Ly$\beta$ and 3\% for Ly$\gamma$,

\begin{subequations}
\begin{align}
\tau_{{\rm eff,} \beta}/\tau_{{\rm eff,} \alpha} &= 0.190 \times (1+z)^{0.434} \label{eqn:ratios_b}\\
\tau_{{\rm eff,} \gamma}/\tau_{{\rm eff,} \alpha} &= 0.0663 \times (1+z)^{0.707} \label{eqn:ratios_g}
\end{align}
\end{subequations}

Assuming a constant ratio over this redshift range is not a bad approximation for Ly$\beta$, with deviations of $<$7\% from the specific simulation values, but there is some systematic bias in doing this.
We therefore use the redshift-dependent version, equation \ref{eqn:ratios_b}, in our helium $\tau_{{\rm eff,} \alpha}$ estimates in Table \ref{tab:alpha_ests}.
The Ly$\gamma$ ratio is constant over this redshift range to within 12\%.
Large-scale structure variations cause a standard deviation over all sightlines of $\simeq$8--10\% in the Ly$\beta$ ratio, and $\simeq$12--19\% in the Ly$\gamma$ ratio, so the ratios could be $\sim$10\% wrong for a specific sightline.
(Although Ly$\gamma$ is not used in the current work, it is included for completeness, as it may be accessible to future observation.
In particular, new modes of COS with higher blue throughput could enable this analysis.)

These results used noise-free, normalized simulated spectra, so to determine if noise has an impact on these results, we create versions that are realistic mock COS spectra.
We put in a power law continuum, redden the spectrum, convolve with the COS LSF, find the expected counts in each pixel, and generate a random realization of each spectrum assuming Poissonian statistics.
We find that noise from random sampling has a negligible effect on our resultant ratio, as one would expect for unbiased noise and a sufficiently large sample.
However, low fluxes and short observing times can affect the observed ratio; e.g., for $f_{1200 \; \mbox{\scriptsize{\AA}}}=5 \times 10^{-17}$~erg~s$^{-1}$~cm$^{-2}$~\AA$^{-1}$ and an exposure time of 3~ks, the mock observation ratio differs from the true simulation ratio by $\simeq$20\%.
However, for all the quasars in this paper, the exposure times and fluxes are sufficient to keep this bias $<$2\%.

While the primary purpose of the simulations is calculating the $z \sim 3$ \ion{He}{2} $\tau_{{\rm eff,} \beta}/\tau_{{\rm eff,} \alpha}$ ratio, we also use them to find that for \ion{H}{1} at $z \sim 0$, $\tau_{{\rm eff,} \beta}/\tau_{{\rm eff,} \alpha} \simeq 0.29$.
We need the latter quantity to remove low-redshift \ion{H}{1} Ly$\beta$ contamination from our \ion{H}{1} Ly$\alpha$ $\tau_{\rm eff}$ measurement, as detailed in Section~\ref{sec:removing_contamination}.

\section{Discussion and Implications for Helium Reionization}
\label{sec:discussion}

We apply the effective optical depth ratios to the Ly$\beta$ measurements to predict helium $\tau_{{\rm eff,} \alpha}$, correcting the Ly$\beta$ optical depths using each of the three methods discussed in section~\ref{sec:removing_contamination}.
The results are presented in Table~\ref{tab:alpha_ests}, the columns of which are as follows.
Column~(1)---target name.
Columns~(2,4,6)---\ion{He}{2}~Ly$\beta$ optical depth measurements, taking the raw optical depths from Table~\ref{tab:obs_opacities} and correcting by methods M1, M2, and M3, respectively.
Columns~(3,5,7)---predicted \ion{He}{2}~Ly$\alpha$ optical depths, using the Ly$\beta$ measurements of Columns~2, 4, and 6, and the theoretical ratios derived in Section~\ref{sec:theory}. Errors quoted are 68\% random only.
Column~8---measurement of the Ly$\alpha$ optical depth in the same redshift range as the Ly$\beta$ measurement. For the SDSS quasars, only 68\% lower limits are quoted, as discussed below.

In some cases, the predicted Ly$\alpha$ optical depth is small enough that we should be able to measure it, allowing us to observationally find $\tau_{{\rm eff,} \beta}/\tau_{{\rm eff,} \alpha}$.
We do this for HE2347, but unfortunately, the SDSS quasars are all contaminated by geocoronal \ion{O}{1}~$\lambda$1302 in the Ly$\alpha$ region corresponding to our Ly$\beta$ measurements.
Although \ion{O}{1} is dramatically weaker at night (Figure~\ref{fig:lya_spec} uses night-only data in that region), it still is strong enough to noticeably bias any measurements in a Gunn-Peterson trough.
Table~\ref{tab:alpha_ests} therefore quotes only 68\% lower limits on the Ly$\alpha$ optical depths for the SDSS quasars, because of this residual geocoronal flux.
The HE2347 measurement and the lower limits on SDSS0915 and SDSS1253 are sufficient to show that M3 appears to be a very poor correction method, substantially underpredicting Ly$\alpha$ opacity.
This is likely due to the large uncertainty in translating $\tau_{{\rm eff, \: H \, I}}$ into $\tau_{{\rm eff, \: He \, II}}$.
Even if the simulations were wrong, and the effective optical depth ratio were as low as possible ($0.16$), the SDSS1253 data strongly disfavor M3.
The measured Ly$\alpha$ values are consistent with the predicted values for M1 and M2, using the simulation ratio.
Given the systematic problems we have seen that M1 has with redshift, we regard M2 as perhaps the best estimate.
This is supported by the good agreement between the M2 prediction and the actual measured Ly$\alpha$ opacity in HE2347 (Table~\ref{tab:alpha_ests}).
In addition, the predictions of M2 are more conservative than those of M1, and therefore, of our three methods, its estimates are the most conservative that fit the data.

Using the M2 predictions, we set a lower limit on the Ly$\alpha$ opacity of $\tau_{{\rm eff, \: He \, II}} > 8.8$ in the SDSS2346 sightline at $\langle z \rangle = 3.38$, and estimate $\tau_{{\rm eff, \: He \, II}} = 4.36^{+1.87}_{-1.15}$ in the SDSS1253 sightline at $\langle z \rangle = 3.35$ and $\tau_{{\rm eff, \: He \, II}} = 3.89^{+3.95}_{-2.00}$ in the SDSS0915 sightline at $\langle z \rangle = 3.29$.
For HE2347, at $\langle z \rangle = 2.78$, we estimate $\tau_{{\rm eff, \: He \, II}} = 2.51^{+0.55}_{-0.49}$.

Our results show that sightline variance, such as that between SDSS1253 and SDSS2346, is quite strong at high redshift, as it was already known to be at low redshift.
It is difficult to turn these mean optical depths into detailed information on the progress of helium reionization, however, in part because of the still-small number of sightlines, but also because theoretical models differ substantially at these redshifts \citep{dixon09,worseck11a}.
Our own simulations were not intended for studying the evolution of helium reionization in detail.
The large sightline variance itself may imply that helium reionization is incomplete at these redshifts \citep{furlanetto10}, but along some sightlines, it was already known to be incomplete at even lower redshifts \citep[e.g., at $z=2.8$ in HE2347;][]{shull10}.
Of course, the large majority of helium ionization and heat injection occur prior to the observational completion of reionization, which can be thought of as the end of black ($\tau \gtrsim 5$) Gunn-Peterson troughs.

The \ion{He}{2} optical depth expectations for a uniform IGM (equation~\ref{eqn:tau_gp}, below) have led to pessimism about being able to observe helium reionization by this method, but these expected values are much higher than they would be in a more realistic inhomogeneous IGM.
\citet{mcquinn09a} discusses some aspects of this, and our simulations show it well in our favored metric, $\tau_{\rm eff}$.
For the WMAP7 parameters \citep{komatsu11} and the helium fraction of \citet{steigman07}, one obtains for a uniform IGM \citep{gunn65}

\begin{align}\label{eqn:tau_gp}
\tau_{{\rm GP, HeII}} &= 3.49 \left( \frac{x_{\rm HeII}}{10^{-3}} \right) \left( \frac{\Delta_b}{1} \right) \left( \frac{Y_P}{0.2486} \right) \left( \frac{\Omega_b}{0.0456} \right) \left( \frac{h}{0.704} \right) \left( \frac{\Omega_m}{0.272} \right)^{-1/2} \left( \frac{1+z}{4} \right)^{3/2} \nonumber \\
& \times \left[ \frac{1.0209}{1+0.0209 \cdot (4/(1+z))^3} \right]
\end{align}

\noindent
where the last term is included to account for the cosmological constant (a small but non-negligible correction at $z=2$--3), $\Delta_b=\rho_b/\langle \rho_b \rangle$ is the baryon overdensity, and $Y_p$ is the helium mass fraction.

For a uniform IGM, therefore, the helium Gunn-Peterson trough becomes saturated to the point that nonzero flux can no longer be measured at $x_{\rm He \, II} \sim 2 \times 10^{-3}$, even for a bright quasar with a long exposure time.
It has been pointed out that in an IGM with density and ionization inhomogeneities, this is no longer the case \citep{furlanetto09a,mcquinn09a}.
Our simulation, while not intended to match the details of helium reionization, provides a clear quantitative example of the effects of density variation.
For example, at an ionization fraction $x_{\rm He \, II} = 2.5 \times 10^{-3}$ (the mass- and volume-weighted ionization fractions $x^M_{\rm He \, II} \simeq x^V_{\rm He \, II}$ in this case), we find helium $\tau_{{\rm eff,} \alpha}=1.65$, rather than the $8.7$ predicted at mean density by equation \ref{eqn:tau_gp}.
The lower value arises because in an inhomogeneous IGM, transmission is dominated by underdense regions.

\section{Conclusion}
\label{sec:conclusion}

We present \ion{He}{2}~Ly$\beta$ Gunn-Peterson trough measurements for three new quasars, and analyze new data on the one quasar for which Ly$\beta$ had been measured before.
The Ly$\beta$ trough has lower opacity than Ly$\alpha$, and thus allows measurements at otherwise unobservable \ion{He}{2} fractions.
We predict large but variable Ly$\alpha$ optical depths at $z \simeq 3.2$--$3.5$, from $\tau_{{\rm eff,}\alpha} \sim 4$ to $>$9.

However, the lower opacity of Ly$\beta$ is not simply predicted by combining the Ly$\alpha$ opacity with the $f_i \lambda_i$ ratio, when dealing with effective optical depths.
We use cosmological simulations to derive $\tau_{{\rm eff,}\beta}/\tau_{{\rm eff,}\alpha} \simeq 0.35$ for helium at $z \sim 3$.
Where both Ly$\beta$ and Ly$\alpha$ measurements are possible, they suggest good agreement between our predictions and the observations, although systematic errors make predicting Ly$\alpha$ from Ly$\beta$ somewhat uncertain.
The \ion{He}{2}~Ly$\gamma$ trough, although a factor of three stronger in $\tau_{\rm eff}$ than in a point-by-point measurement, will be challenging to use.
The short wavelength span of this trough and its lower wavelength position (at lower detector sensitivity) will make the opacity measurements quite difficult, and the corrections will need to include both Ly$\alpha$ and Ly$\beta$, and thus be that much more uncertain.

We present a trough optical depth calculation method suitable for COS (with confidence intervals appropriate for any photon-counting device).
Because of the unified nature of Feldman-Cousins confidence intervals, they allow comparison between all sets of data using this method, regardless of whether the intervals are one- or two-sided.
This aspect is important for any theoretical studies that wish to compare helium reionization simulations to data from different surveys.

\ion{He}{2}~Ly$\beta$ Gunn-Peterson troughs appear to be an effective way to measure high optical depth regions, and new short-wavelength modes of COS ({\it HST}/GO programs 12501 and 12505) will be able to make better measurements of this along several sightlines.

\acknowledgments

We thank Tae-Sun Kim for providing the continuum-normalized VLT spectrum of HE2347.

Support for {\it HST} Programs number 12178 and 12249 was provided by NASA through grants from the Space Telescope Science Institute, which is operated by the Association of Universities for Research in Astronomy, Incorporated, under NASA contract NAS5-26555.

\begin{deluxetable}{lllr}
\tablecolumns{9}
\tablewidth{0pc}
\tablecaption{Observations}
\tablehead{
\colhead{Target} & \colhead{Grating\tablenotemark{a}} & \colhead{Obs. Date} & \colhead{Exp. Time} \\
\colhead{} & \colhead{} & \colhead{} & \colhead{(s)}}
\startdata
HE2347-4342  & G130M/G140L & 2009 Nov 5 & 28,458/11,558 \\
SDSSJ0915+4756 & G140L & 2010 Oct 6 & 5,521 \\
SDSSJ1253+6817 & G140L & 2011 May 5 & 14,096 \\
SDSSJ2346-0016 & G140L & 2010 Nov 29, Dec 4 & 20,737 \\
\enddata
\tablenotetext{a}{Of {\it HST}/COS. G130M covers $\lambda=1134$--1474~\AA, while G140L covers (at greatly varying sensitivity) $\lambda \sim 920$--2000~\AA, where the lower limit is from the Galactic Lyman limit. Both gratings have multiple central wavelength settings that shift coverage, and the values quoted are extremes.}
\label{tab:obs}
\end{deluxetable}

\begin{deluxetable}{lcccccc}
\rotate
\tablecolumns{9}
\tablewidth{0pc}
\tablecaption{Measured \ion{He}{2}~Ly$\beta$ Optical Depths}
\tablehead{
\colhead{Target} & \colhead{Quasar Redshift} & \colhead{\ion{He}{2}~Ly$\beta$ redshift range} & \colhead{Raw $\tau_{{\rm eff,} \beta}$} & \colhead{M1 $\tau_{{\rm eff,} \alpha}$} & \colhead{M2 $\tau_{{\rm eff,} \alpha}$} & \colhead{M3 $\tau_{{\rm eff,} \alpha}$}}
\startdata
HE2347-4342  & 2.887 & 2.684--2.880 & $1.75^{+0.14}_{-0.12}$ & 1.07 & $0.92^{+0.11}_{-0.10}$ & $1.374 \pm 0.044$ \\
SDSSJ0915+4756 & 3.343 & 3.253--3.329 & $3.46^{+1.37}_{-0.69}$ & 1.66 & $2.10^{+0.12}_{-0.11}$ & $2.54 \pm 0.40$ \\
SDSSJ1253+6817 & 3.470 & 3.292--3.414 & $3.99^{+0.66}_{-0.40}$ & 1.74 & $2.433^{+0.084}_{-0.080}$ & $3.66 \pm 0.52$ \\
SDSSJ2346-0016 & 3.510 & 3.276--3.493 & \phd$[6.02,\infty)\tablenotemark{a}$ & 1.79 & $2.814^{+0.070}_{-0.068}$ & $3.24 \pm 0.34$ \\
\enddata
\tablecomments{Method 1 (M1) uses an average semianalytical approximation to $\tau_{{\rm eff,} \alpha}(z)$, method 2 uses $\tau_{{\rm eff,} \alpha}$ from just above the Ly$\beta$ break, and method 3 uses the \ion{H}{1}~Ly$\alpha$ optical depth to predict $\tau_{{\rm eff,} \alpha}$. See Section \ref{sec:removing_contamination} for further details.}
\tablenotetext{a}{No finite best estimate exists for this quantity. The sensitivity is $\tau_{\rm s}=5.59$ (68\%), and the detector lower limit is $\tau_{\rm dll}=5.15$ ($\alpha=0.32$, $\beta_{\rm min}=0.9$). (See Section \ref{sec:tau_methods} of the text for discussion of sensitivity and detector lower limit.) Using the night spectrum only, we obtain $\tau_{{\rm eff,}\beta}=6.22^{+2.22}_{-0.99}$.}
\label{tab:obs_opacities}
\end{deluxetable}

\begin{deluxetable}{lrccc}
\tablecolumns{9}
\tablewidth{0pc}
\tablecaption{Simulation Predictions for the Optical Depth Ratio}
\tablehead{
\colhead{Transition} & \colhead{Wavelength} & \colhead{Oscillator Strength ($f$)} & \colhead{$\tau_{i}(\lambda)/\tau_{\alpha}(\lambda)$} & \colhead{$\tau_{{\rm eff,} i}/\tau_{{\rm eff,} \alpha}$\tablenotemark{a}}}
\startdata
Ly$\alpha$ & $303.782$ & $0.4162$\phn & $1$\phn\phn\phn\phn\phd & $1$\phn\phn\phd \\
Ly$\beta$ & $256.317$ & $0.07910$ & $0.160$\phn & $0.35$ \\
Ly$\gamma$ & $243.027$ & $0.02899$ & $0.0557$ & $0.18$ \\
\enddata
\tablenotetext{a}{As found in simulations for $z=3.3$; see Section \ref{sec:theory}. These do not vary strongly with redshift.}
\label{tab:theory_opacities}
\end{deluxetable}

\begin{deluxetable}{lccccccc}
\rotate
\tablecolumns{9}
\tablewidth{0pc}
\tabletypesize{\footnotesize}
\tablecaption{Predicted \ion{He}{2}~Ly$\alpha$ Optical Depths}
\tablehead{
\colhead{Target} & \colhead{$\tau_{{\rm eff,} \beta}$, corr.\ M1} & \colhead{$\tau_{{\rm eff,} \alpha}$, pred.\ M1} & \colhead{$\tau_{{\rm eff,} \beta}$, corr.\ M2} & \colhead{$\tau_{{\rm eff,} \alpha}$, pred.\ M2} & \colhead{$\tau_{{\rm eff,} \beta}$, corr.\ M3} & \colhead{$\tau_{{\rm eff,} \alpha}$, pred.\ M3} & \colhead{$\tau_{{\rm eff,} \alpha}$, measured\tablenotemark{a}}}
\startdata
HE2347-4342 & $0.68^{+0.14}_{-0.12}$ & $2.06^{+0.43}_{-0.37}$ & $0.83^{+0.18}_{-0.16}$ & $2.51^{+0.55}_{-0.49}$ & $0.38^{+0.15}_{-0.13}$ & $1.14^{+0.45}_{-0.39}$ & $2.393 \pm 0.015$ \\
SDSSJ0915+4756 & $1.80^{+1.37}_{-0.69}$ & $5.14^{+3.91}_{-1.97}$ & $1.36^{+1.38}_{-0.70}$ & $3.89^{+3.95}_{-2.00}$ & $0.92^{+1.43}_{-0.80}$ & $2.63^{+4.09}_{-2.29}$ & $>3.84$ \\
SDSSJ1253+6817 & $2.25^{+0.66}_{-0.40}$ & $6.25^{+1.84}_{-1.12}$ & $1.56^{+0.67}_{-0.41}$ & $4.36^{+1.87}_{-1.15}$ & $0.33^{+0.84}_{-0.33}$ & $0.92^{+2.34}_{-0.92}$ & $>3.55$ \\
SDSSJ2346-0016 & $[4.23,\infty)$ & $[11.59,\infty)$ & $[3.16,\infty)$ & $[8.78,\infty)$ & $[2.54,\infty)$ & $[7.06,\infty)$\phn & $>4.49$ \\
\enddata
\tablecomments{~Ly$\beta$ optical depths are corrected using the estimates of Table~\ref{tab:obs_opacities}, and Ly$\alpha$ optical depths are predicted using the ratio of Section~\ref{sec:theory}. Errors quoted are random only.} 
\tablenotetext{a}{Only lower limits exist for the three SDSS quasars, because of contaminating geocoronal \ion{O}{1} emission.}
\label{tab:alpha_ests}
\end{deluxetable}

\end{document}